\documentstyle[prl,aps]{revtex}
\begin{document}
\twocolumn
\draft
\title { The Addition Spectrum of a Lateral Dot from Coulomb and Spin Blockade Spectroscopy}

\author{M.\ Ciorga$^{1,2,\dag}$, A. S. Sachrajda$^{1}$,
P.Hawrylak$^{1}$, C. Gould
$^{1,2}$,  P. Zawadzki$^{1}$, S.Jullian$^{3}$, Y. Feng$^{1}$, and Z.
Wasilewski$^{1}$\\ }

\address{
\
\protect${^1}$Institute for Microstructural Science, National Research
Council, Ottawa K1A 0R6, Canada\\
\protect$^2$ D\'{e}partement de Physique and CRPS, Universit\'{e} de
Sherbrooke,Sherbrooke J1K 2R1, Canada\\
\protect$^3$Grenoble High Magnetic Field Laboratory, MPI/FKF and CNRS\\
25, av des Martyrs, F-38042 Grenoble Cedex 9, France\\}
 
\date{\today}
 
\wideabs{ 
\maketitle

\begin{abstract}
Transport measurements are presented on a
class of electrostatically defined
lateral dots within a high mobility two
dimensional electron gas (2DEG). The new design allows Coulomb Blockade(CB)
measurements to be performed on a single lateral dot containing 0,
1 to over 50 electrons. The CB measurements are enhanced by 
the spin polarized injection from and into  2DEG magnetic edge states. This combines the measurement
of charge with the measurement of spin  
 through spin blockade spectroscopy.
The results of Coulomb and spin blockade spectroscopy
for first 45 electrons enable us to construct the addition spectrum
of a lateral device. We also demonstrate that a lateral dot containing a single electron
 is an effective local probe of a 2DEG edge.
 
\end{abstract}

\
}
 
The ability to confine, manipulate, and probe individual electrons in a solid state
environment is a prerequisite for a variety of transport nano-scale applications, from
single electron transistors to gates in a quantum computer\cite{dot reviews,loss,pawel cmn}.
 It is an important task therefore to construct a quantum dot in which one can simultaneously
tune the number of electrons and interrogate them spectroscopically.
This has been achieved previously in vertical quantum dots in which the number of confined electrons has been successfully
controlled down to zero and a number of spectacular new discoveries made using both capacitance and Coulomb blockade 
techniques\cite{tarucha,ashoori}.
Vertical devices are pillars formed by
etching out material in a double barrier tunneling device.
The electrons are injected into the pillar from n$^+$ contacts.
Early pioneering investigations have also been performed on lateral devices \cite{mceuen}. While the versatility of 
electrostatically defined lateral dots formed within a high mobility two dimensional electron gas(2DEG) was well
established, it has always been assumed to be impossible to empty a lateral dot while
maintaining operating tunneling barriers \cite{leo}. It has also proved difficult to extend measurements to very
high magnetic fields as a result of pinch off effects associated with magnetic depopulation within the tunneling barriers.
 Recently we
 managed to reduce the number of electrons in a lateral dot to around 10 \cite{gould1,pawel andy prb 1999}. It was
 discovered in those 
measurements that above a certain magnetic field, electrons entering the dot were spin polarized.
 This was a direct consequence of their injection from spin polarized magnetic edge states in the 2DEG which 
formed at the tunneling barriers into and out of the dot. 
We were able to make use of this additional 'spin spectroscopy' to confirm the role that spin textures play in 
the spin-flip regime (2$>$$\nu$$>$1 where $\nu$ is the filling factor) \cite{pawel andy prb 1999,gould}.

  In this paper we demonstrate that
lateral dots defined in a 2DEG can be completely
and controllably emptied of
electrons. We are able not only to measure the addition spectrum of a lateral dot 
but are able to extend the 'spin' spectroscopy to this lower field regime (i.e. $\nu$$>$2)
and $\it{directly}$ observe novel quantum dot spin phenomena. Previously spin related physics such as Hunds rule 
and single-triplet transitions were indirectly deduced from the addition spectrum. Finally we discuss how the
 quantum dot can be used 
as a local probe of the 2DEG edge.

The lower inset of figure 1 is a SEM micrograph of a device similar to the one used in our
measurements.
The dot is electrostatically defined by four gates above a GaAs/AlGaAs
heterostructure.  A
top 'T' gate in combination with left and right finger gates is used
to define the dot geometry. A narrow plunger gate, located in the gap between the left and right finger gates 
is used to vary the
 number of electrons.
 The lithographic width and height of the triangular dots was approximately
$0.45 \mu$m. The bulk density (mobility) of the
AlGaAs/GaAs wafer used were 1.7$\times 10^{11}$cm$^{-2} (2\times
10^{6}$m$^{2}$V$^{-1}$ s$^{-1}$). To operate the device in the Coulomb blockade
mode the top and finger gates are
 used to create tunneling barriers between the 2DEG and the dot. The plunger gate is
used to vary the number of electrons in the dot. A Coulomb blockade peak
is observed whenever the electron number
in the dot changes. As was first shown by McEuen et al.\cite{mceuen}, the gate voltage
position of the Coulomb blockade peaks is related not only to the charging energy 
but also provides 
information about the kinetic energy  spectrum of the dot. The shape, gate design and MBE wafer were chosen to enable us to
reach the minimum number of electrons
in the dot while still enjoying operating tunnel barriers. In particular, it is crucial
that two gates define each tunneling barrier, one with large and one with small
contributions to the dot perimeter. 
This allowed several operations to be achieved that were critical to (i) the
successful emptying of the lateral dot and (ii) for extending spectroscopic measurements to high magnetic fields (18T).
Applying a more negative voltage to the finger
 gates while keeping the same tunnel barrier height (achieved by simultaneously
reducing the voltage applied to the T gate)
 reduced the size of the dot. Magnetic depopulation effects in the barrier
region were compensated for by miniscule reductions
(millivolts) in the voltage applied to the T gate.

Figure 1 plots the first 22 peaks at low magnetic fields, from 0 to 2T. A series of sweeps is made
using the techniques described above to systematically
reduce the size of the dot and number of electrons. Overlapping ranges are chosen to ensure
that the changes made are sufficiently
adiabatic that identical behavior is observed for the peaks present in overlapping sweeps. The gate voltage
scale in figure 1 was set by
normalizing the overlapping curves at 1T. In any device with soft confinement
the single particle energy spectrum
$E(m,n)=\Omega_-( m+1/2)+\Omega_+( n+1/2)$
is that of a pair of harmonic oscillators with magnetic field tunable
frequencies $\Omega_{\pm}$, the Fock-Darwin (FD) spectrum.\cite{fockdarwin,pawel secs}
 At zero magnetic field electronic shells of degenerate
levels exist. Weakly interacting electrons fill up these shells according to
Hund's rules. The complete shell filling is expected for
N=2,6,12,20,.. electrons. When a shell is filled, it costs extra kinetic
 energy to add an electron leading to an increased peak spacing. The inset in figure 1 plots the zero field peak 
spacing betweeen the N and N+1 peaks against N. The spacing between peaks
generally increases as the number of electrons and the size of the dot is reduced. This is particularly apparent
for the first few electrons. The arrows mark the electron number corresponding to the expected full shells
 \cite {tarucha}. Agreement with shell effects (i.e. increased peak spacing) is apparent only at 6 
electrons. Consequences of the dot rapidly shrinking may mask an increased peak level spacing at 2 electrons but the spacing
after 12 and 
20th electrons shows 
no evidence of shell structure. 
A simple interpretation in terms of shells seems unlikely. We speculate that this is related to the reduced screening 
conditions in lateral dots which might be expected to result in a more strongly interacting electron system. 
There is in addition no obvious spin-related odd-even
behavior in peak spacing at zero field such as might be expected if the confinement energy dominated over the 
interaction terms.  

The first 45 peaks in the same magnetic field range as figure 1
but with the charging energy removed by
shifting the peaks together are illustrated in figure 2. The CB peak positions plot out the addition spectrum for a dot with 
a fixed number of 
electrons. The addition spectrum reflects the kinetic energy of the added electron. In the simplest model of a 
parabolic dot, the magnetic field
induces level crossings of single particle eigenstates.
 Level crossings appear as cusps in the peak
positions.
The oscillations in peak positions associated with the level crossings terminate
at $\nu=2$ boundary. There are several
experimental confirmations that we have fully emptied the dot. Firstly, no further CB peaks
are found using the
systematic techniques similar to those described above. Secondly, the experimental curves of figure 2 clearly show the
 $\nu$=2 line extrapolating completely to zero magnetic field as expected for
an emptying dot and as observed in the vertical dot experiments. Thirdly, we can extract the number 
electrons from the number of spin flips from  $\nu$=2 to $\nu$=1
dot\cite{pawel andy prb 1999}.
 Clearly one needs to flip half the spins to polarize an unpolarised dot.
 We can confirm, based on results from these new dots, that this technique is consistent
with the electron number obtained by counting peaks upwards from an empty dot.  
 
        We now concentrate on two important features of these measurements that are a direct consequence of using 2DEG 
leads rather than n$^{+}$ contacts.

The benefits of spin polarized injection for spectroscopic purposes reveals itself in the amplitude of the Coulomb blockade
 peaks. Spin effects can be 'directly' observed by current readout (i.e. by monitoring the effects of spin blockade on the 
Coulomb blockade peak amplitude).
 The inset in figure 3 plots the peak amplitude in the 
Fock-Darwin regime of the addition spectrum for several peaks. Starting around 0.4T (the same magnetic field at which the 
spin resolution of the 2DEG Shubnikov-de Haas oscillations is first seen in this GaAs/AlGaAs wafer) a digital amplitude 
behavior can be observed in the amplitude spectrum. We expect a reduced peak amplitude if the difference between the N and
 N+1 electron groundstates is either (i) a spin up electron ( since only spin down electrons are injected from the 
innermost edge state) or (ii) an electron near the center of the dot (due to a reduced wavefunction overlap with the
 electrons at the edge). Figure 3 relates the amplitude to the peak position spectrum for a few typical 'digital' 
features as shown in the inset. A pattern is seen to repeat itself. Downward sloping lines of roughly equal length A-B and
 C-D correspond to adding the incoming electron to (m,0)and (m+1,0) levels (i.e. from the lowest Landau level) at 
the edge of the dot. The short (B-C) and long (D-A) vertical steps both correspond to adding an electron to higher Landau 
level eigenstates closer to the center of the dot. The short(long) vertical step occurs at a large(small) to small(large) 
peak amplitude transition for the downward sloping lines. Since the A-B and C-D regions correspond to adding an electron 
at the dot edge their amplitude simply reflects whether the difference between the N and N+1 electron groundstates is
 a spin up 
or down electron. It is important to note that this alternating large/small amplitude behavior is not consistent 
within a non-interacting picture in which the downward sloping regions would either have a small(spin up) or a large 
(spin down) amplitude but not alternating large/small amplitudes along a single peak (nb. Zeeman splitting can be 
ignored at these low fields). We have observed identical behavior for lower electron numbers at the less
 complex theoretically $\nu$=2 boundary where the effect can be accurately attributed to an interaction related 
singlet-triplet 
transition \cite{wojs}(further details of these spin effects at $\nu$=2 and a comparison with calculations 
will be published elsewhere). Tarucha et al.\cite{tarucha1} have also deduced the 
singlet-triplet transitions at $\nu$=2 using the position of CB peaks
in the addition 
spectrum alone. The similarity with the $\nu$=2 regime suggests that similar interaction driven spin effects also occur at much 
lower fields. By analogy with the theory at $\nu$=2 the data is consistent with following description of the magnetic field
evolution of even number ground states 
in the Fock-Darwin regime. As the magnetic field is lowered electrons at the edge are
transferred one by one to higher 
harmonic levels near the center of the dot. Each of kinetic energy levels at the edge contains a spin down and a spin 
up electron. The spin up electron is transferred first as the field is lowered but due to interactions it is energetically
favorable
for it to flip spin and form a triplet state with its former partner. As the field is lowered further the singlet pairing is 
reinstated  as the second electron is 
transferred to the same state as the first but as a spin up electron. The cycle is
 then repeated. The additional dip in amplitude during the D-A step occurs because the incoming electron in that region is 
being added to the center of the dot.

Secondly let us consider the effect of the 2DEG chemical potential jumps associated with Landau level 
depopulation on our measurements. While most cusps in figure 2 are peak number dependent, certain features are 
identical for all peaks. These
can be seen most clearly for N=1 peak where we do not expect or see any features related to level crossings. 
Figure 4 shows this peak from 
B=-0.5 to 3T. To understand the origin of the step like
structures observed on this Coulomb
blockade peak it is important to remember that the Coulomb blockade occurs when
the chemical potentials of the dot and
leads are matched. The 1/B step like features are due to the chemical potential jumps in
the 2DEG. The $\nu$=1 and
$\nu$=2 2DEG chemical potential jumps
at higher fields have also been observed. The inset in figure 4 shows a schematic of the bulk chemical potential of a 2DEG.
 We note that with the exception of $\nu$ =1 only the even filling factor steps are resolved experimentally. The 2DEG
 chemical potential steps are observed in an inverted fashion in the data since a drop in the 2DEG chemical potential requires
a less negative plunger gate voltage to maintain the resonance condition. The first peak (i.e. that observed 
on adding the first electron) is crucial for spectroscopic investigations since it can be used to separate 2DEG effects 
which will be present on this peak from intrinsic quantum dot effects which will not. The observations shown in figure 4 
also confirm that a lateral dot containing a single electron can be used as a local probe of the 2DEG edge chemical 
potential \cite{wei}.

In conclusion, we have demonstrated  two important 
 features of  the lateral nature of a quantum dot which emerge when we are capable of emptying the dot
and also able to perfoem transport measurements.  
We have demonstrated that even in the Fock-Darwin regime of the addition spectrum the use of a high mobility 2DEG 
leads to spin polarized injection into the dot. This provides a new spectroscopic tool, the spin blockade,
which directly measures the spin 
state of the dot through 'current readout' and allows novel spin phenomena to be identified in the addition spectrum. 
The second feature is the application of an empty quantum dot as a spectrometer of edge states of 2DEG.
The first, N=1, peak may be used for locally probing the 2DEG edge and these studies,
 interesting in themselves, can then be utilized to separate quantum dot features for higher electron numbers. 
Finally we note that isolation of a single spin and its probing with a spin polarized source of electrons constitutes
 readout of a spin qubit in a quantum computer.

$\dag$ Permanent Address: Institute of Physics, Wroclaw University of Technology, 
Wybrzeze Wyspianskiego 27, 50-370 Wroclaw, Poland.

\begin{figure} \caption{  A plot of the first 22 Coulomb blockade peaks. The
arrows refer to the expected
full shells for a circular dot. The upper inset shows the peak spacing. The lower
inset is a SEM of a device similar to the one used in the measurements.}
\label{fig1} \end{figure}

\begin{figure} \caption{ The first 45 peaks with the charging energy manually
removed. The arrow points to the $\nu$=2
line which extrapolates to zero magnetic field as we empty the dot.}
\label{fig2} \end{figure}

\begin{figure} \caption{A comparison of the peak position and amplitude behavior at low fields for peak 43. 
The inset shows the low field $\it{amplitude}$ spectrum of peaks
(36 to 45).} 
\label{fig4} \end{figure}

\begin{figure} \caption{The data for the first peak. The step like structures
are a consequence of the
chemical potential jumps in the 2DEG. The inset is a schematic of the 2DEG chemical potential.
 }  \label{fig3} \end{figure}


\begin{references}
\bibitem{dot reviews}
For reviews and references see L. Jacak, P. Hawrylak, and A. Wojs, {\it Quantum
Dots}, Springer Verlag Berlin, 1998; 
R. C. Ashoori, Nature {\bf 379},413 (1996),
M.Kastner, Physics Today,{\bf 44}, 24 (1993);
T. Chakraborty, Comments in Cond.Matter Physics {\bf 16},35(1992).

\bibitem{loss}
D. Loss, D. P. DiVincenzo Phys. Rev.A 57, 120 (1998).

\bibitem{pawel cmn}
P. Hawrylak, S. Fafard, and Z. Wasilewski
Condensed Matter News {\bf 7}, 16 (1999).

\bibitem{tarucha} S. Tarucha, D. G. Austing, T. Honda, R. J. van der Hage,
and L. P. Kouwenhoven, Phys. Rev. Lett {\bf 77}, 3613 (1996).



\bibitem{ashoori} 
R. C. Ashoori, H. L. Stormer, J. S. Weiner, L. N. Pfeiffer,
K. W. Baldwin, and K. W. West, Phys. Rev. Lett. {\bf 71}, 613 (1993).


\bibitem{mceuen} 
P. L. McEuen, E. B. Foxman, J. M. Kinaret, U. Meirav, M. A. Kastner,
N. S. Wingreen, and S. J. Wind, Phys. Rev.{\bf B45}, 11 419 (1992).
O. Klein, S. de Chamon, D. Tang, D. M. Abusch-Magder, U. Meirav,
X.-G. Wen, M. A. Kastner, and S. J. Wind,
Phys. Rev. Lett. {\bf 74}, 785 (1995).

\bibitem{leo}
L.P. Kouvenhoven, C.M. Marcus, P.L. McEuen, S. Tarucha, R.M. Westervelt, and N.S. Wingreen
Section 5.1, Nato ASI conference proceedings, ed. By L. P. Kouwenhoven, G. Schön, L.L. Sohn (Kluwer, Dordrecht, 1997).



\bibitem{gould1}
C. Gould, P. Hawrylak, A. S. Sachrajda, Y. Feng, Z. Wasilewski
Physica {\bf B 256}, 141 (1998).

\bibitem{pawel andy prb 1999}
P. Hawrylak, C. Gould, A. Sachrajda, Y. Feng, Z. Wasilewski,
Phys. Rev. {\bf B 59}, 2801(1999).


\bibitem{gould} 
C. Gould {\it et al.} Physica E {\bf 6}, (2000) 461.

\bibitem{fockdarwin} V. Fock, Z. Phys. {\bf 47}, 446 (1928); C. G. Darwin,
Proc. Cambridge Philos. Soc. {\bf 27}, 86 (1930).

\bibitem{pawel secs}
P. Hawrylak, Phys. Rev. Lett. {\bf 71}, 3347 (1993).


\bibitem{wojs}
A.Wojs and P.Hawrylak, Phys. Rev. {\bf 53} 10841 (1996).

\bibitem{tarucha1}
S. Tarucha, D.G. Austing, Y. Tokura, W.G. van der Wiel, L.P. Kouwenhoven
Phys. Rev. Lett.{ \bf 84}, 2485 (2000).

\bibitem{wei}
Wei. et al.  have used a metallic single electron transistor to study the chemical potential
 variations of a 2DEG. Y.Y.Wei, J.Weis, K. v. Klitzing and K.Eberl Appl. Phys. Rev. Lett. {\bf 71} 2514 (1997).










\end{references}
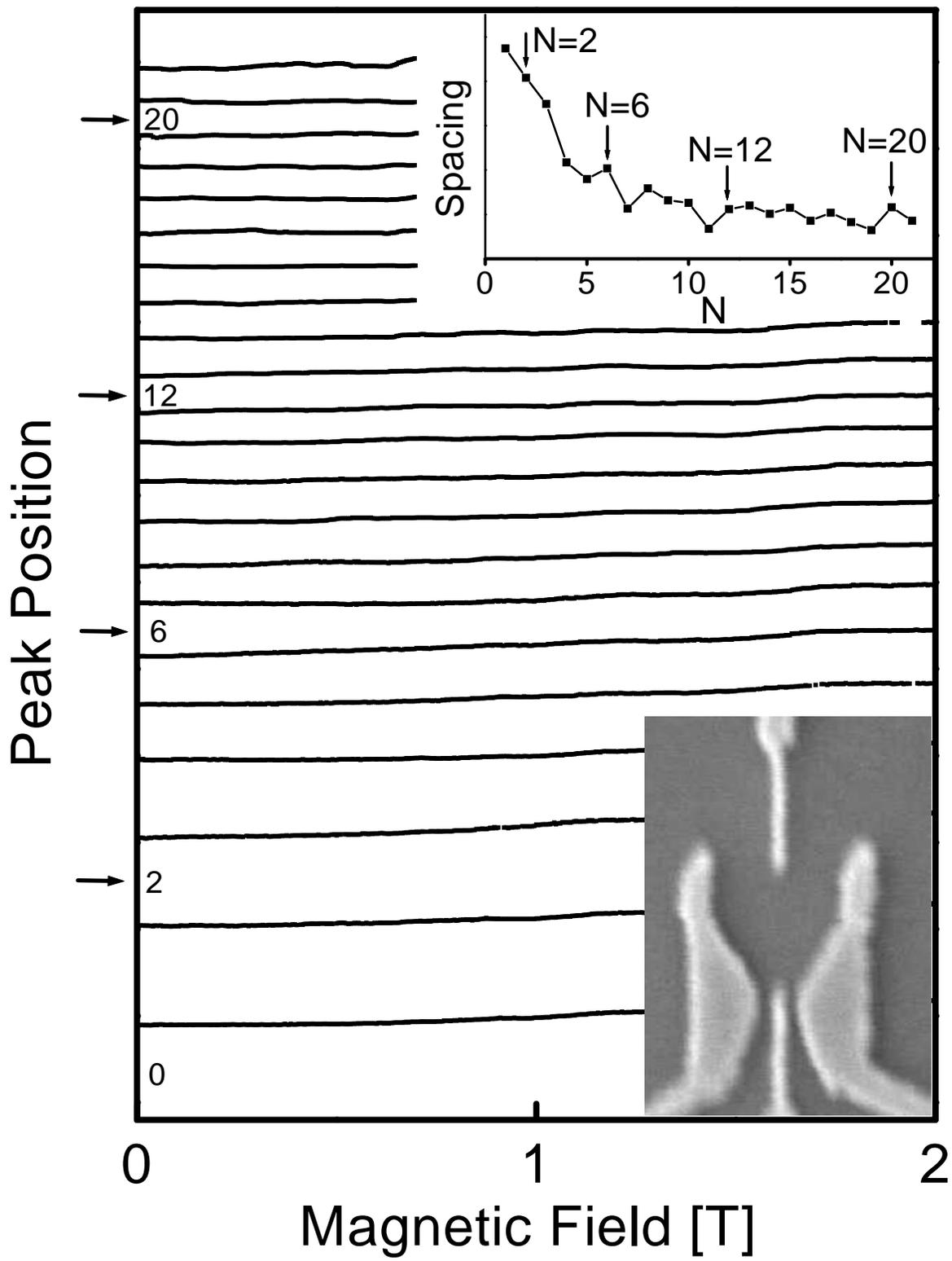
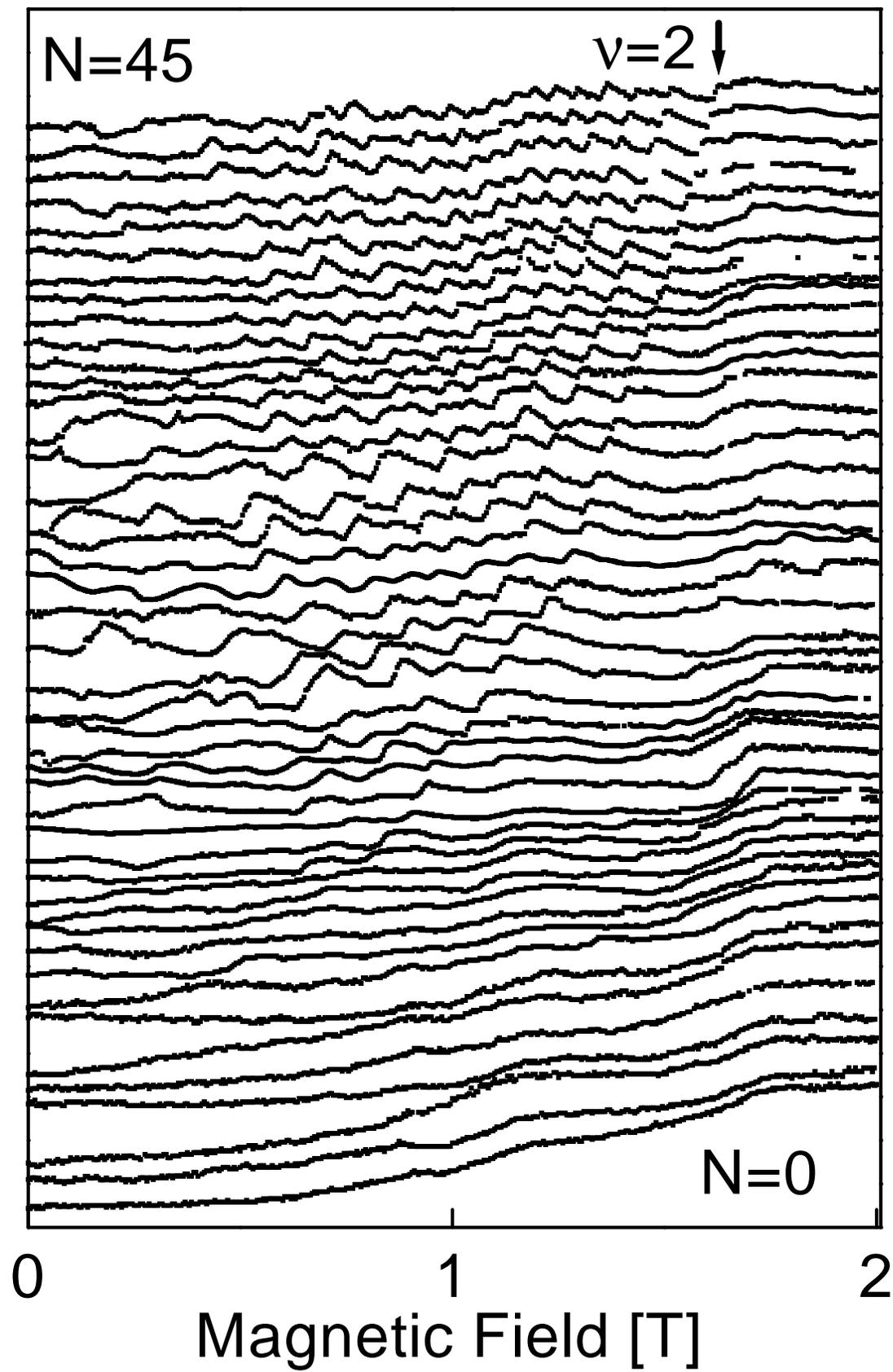
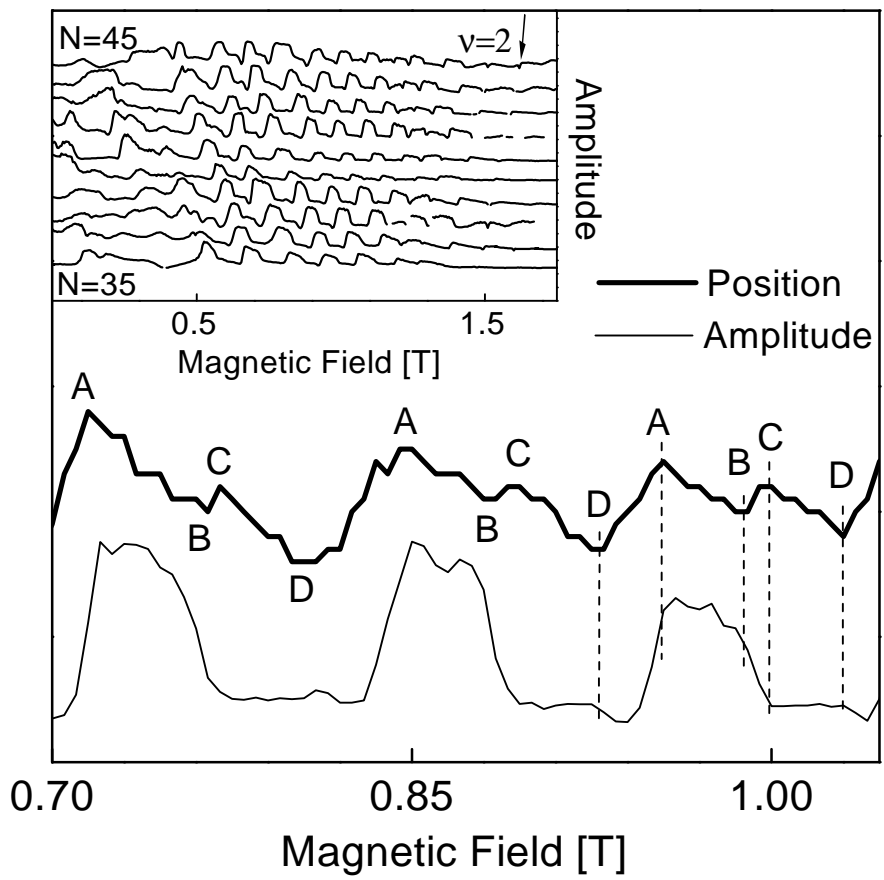
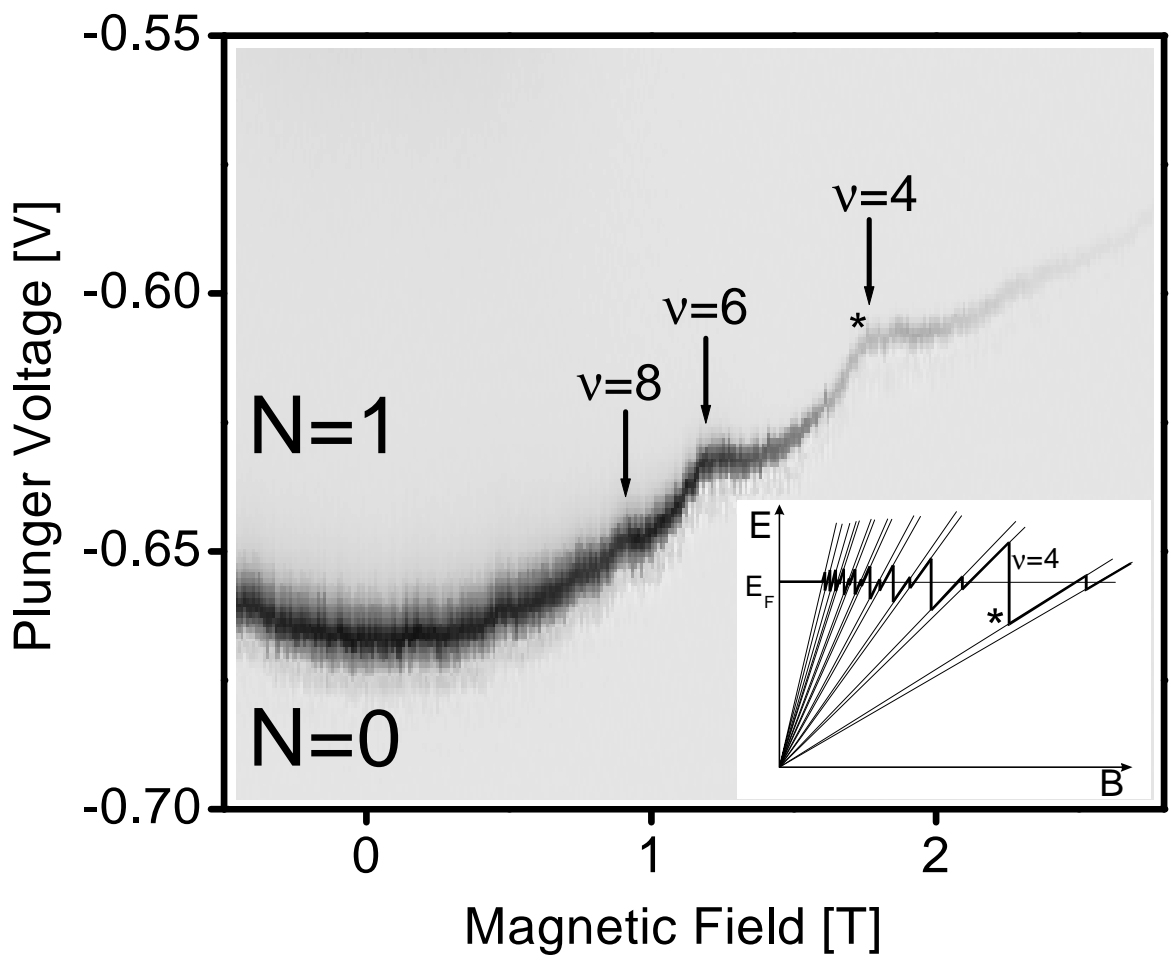
\end{document}